\documentclass[prx,onecolumn,nofootinbib,citeautoscript,eqsecnum,10pt,longbibliography,notitlepage]{revtex4-2}

\pdfoutput=1
\usepackage{booktabs,makecell}
\usepackage{graphicx,amsmath,amssymb,braket,bbm,bm,mathtools}
\usepackage{latexsym}
\usepackage{color,braket}
\usepackage{titlesec}
\usepackage{dsfont}
\usepackage[tight]{subfigure} 
\usepackage[papersize={8.5in,11in}]{geometry}
\usepackage[pdftex,colorlinks=true,linkcolor=darkblue,citecolor=blue,urlcolor=darkred]{hyperref}

\definecolor{darkblue}{rgb}{0.,0.,0.4}
\definecolor{darkred}{rgb}{0.5,0.,0.}
\definecolor{BlueViolet}{RGB}{138,43,226}
\definecolor{SkyBlue}{RGB}{30,144,255}
\definecolor{DarkGreen}{RGB}{0,100,0}

\def\*#1{\mathbf{#1}} 
\def\mrm{\mathrm}

\begin{document}
\title{Magnus Hall effect in three-dimensional topological semimetals}

\author{Sajid Sekh}
\affiliation{Institute of Nuclear Physics, Polish Academy of Sciences, 31-342 Krak\'{o}w, Poland}

\author{Ipsita Mandal}
\affiliation{Institute of Nuclear Physics, Polish Academy of Sciences, 31-342 Krak\'{o}w, Poland}

\begin{abstract} 
Magnus Hall effect (MHE) is a non-linear Hall effect requiring no external magnetic field, which can be observed when an in-built electric field couples to the Berry curvature of the bandstructure, producing a current in the transverse direction. In this paper, we explore MHE in the context of various three-dimensional semimetals, incorporating various features like tilt, anisotropy, and multi-fold degeneracy. We numerically calculate the Magnus Hall conductivity tensors and transport coefficients, within the framework of the Boltzmann transport theory. Although MHE was originally predicted for two-dimensional materials with time-reversal symmetry (TRS), we show that a finite MHE response is possible in materials without TRS. If TRS is preserved, broken inversion symmetry is needed to prevent the cancellation of the MHE contributions while summing over the Brillouin zone.
The amount of tilt of the node of a semimetal greatly affects the transport coefficients. In presence of anisotropic dispersions, we find that the MHE features differ depending on the directions of measurements (as expected). To demonstrate these dependencies, our investigations include Weyl, multi-Weyl, multi-fold, and nodal-line semimetals. Our analysis is of great importance for transport measurements in experiments involving non-linear Hall effects.
\end{abstract}

\maketitle

\tableofcontents
\section{Introduction}

An electron, moving under the combined influence of an in-built electric field (inside a material) and a perpendicular magnetic field, starts to drift due to the Lorentz force, thus generating a current in the direction perpendicular to both the electric and magnetic fields. Discovered by Edwin Hall, this interplay of electric and magnetic field leads to the much-celebrated ``classical Hall effect''. In a prototypical setup, the Hall voltage is directly proportional to the applied magnetic field, which is typically equal to a few units of Tesla. If the strength of the magnetic field is further raised to $\sim 100$ T, the Hall conductance becomes quantized, leading to integer~\cite{klitzing} and fractional~\cite{stormerFQH} quantum Hall effects. In this regime, several interesting phenomena are expected to occur, such as chiral edge states~\cite{patlatiuk} and charge fractionalization~\cite{horsdal}. 
These observations make the external magnetic field an essential ingredient to observe the Hall effect. But recent scientific endeavours suggest that Hall effects without a magnetic field are also possible, mainly through two ways: (1) by adding spin or orbital degrees of freedom; (2) going beyond the linear response regime. The former route leads to quantum spin Hall effect~\cite{bernevig} in a topological insulator (TI), where spin-orbit coupling (SOC) acts as an artificial gauge field. The latter avenue refers to the non-linear Hall effects, which appear even when time-reversal symmetry (TRS, denoted by $\mathcal{T}$) is preserved, but require a broken inversion symmetry (denoted by $\mathcal{I}$). Mathematically, if the transverse Hall current $\mathbf J $ is expanded in terms of the generating electric field $\mathbf E $, then the first-order (linear in the components $\mathbf E$) terms $\sigma_{ab}\,E_b$ refer to the linear Hall effect, while the non-linear effect arises from the generation of currents as dictated by the second-order terms of the form $\chi_{abc} \, E_b \, E_c$~\cite{du,pacchioni}. To measure the current experimentally, an alternating current of low frequency is injected into the sample, which generates an oscillating voltage at `double-frequency' in the case of non-linear Hall effect -- a feature that is not present in its linear counterpart. Depending on the driving current (electrical, spin, or thermal) involved, non-linear Hall effects can be categorized into different types, namely non-linear spin Hall effect~\cite{hamamoto}, gyrotropic effect~\cite{konig}, Magnus Hall effect (MHE)~\cite{papaj,agarwal,das}, and non-linear Nernst effect~\cite{zeng,yu}. Several first-principle calculations predict such non-linear effects in the transition metal dichalcogenides~\cite{low, du2}, crystalline topological insulators~\cite{sodemann}, and Weyl semimetals~\cite{facio,zhang}. In addition, experimental observations have been made in two-dimensional (2D) layers of $\mathrm{WTe}_2$~\cite{kang} and non-magnetic topological insulators $\mathrm{Bi}_2\mathrm{Se}_3$~\cite{he}.

In this paper, we focus on the MHE, which is a Berry curvature (BC) induced Hall effect, requiring no external magnetic field. Classically, the Magnus effect is observed when a spinning object experiences a change in its trajectory while moving through a fluid, such that its path is deflected in a manner that is absent when the object is not spinning. The MHE is the condensed matter analogue of the above, where chiral Bloch electrons, under an electrostatic potential gradient, develop a velocity in the transverse direction (cf. Fig.~\ref{fig:cartoon}). The two main ingredients needed for MHE are: (1) a slowly varying potential gradient, which generates an in-built electrical field; and (2) a non-zero BC in the Brillouin zone (BZ). The system needs to have either broken-$\mathcal T$ or broken-$\mathcal{I}$ in order to have a non-zero BC, which drives the MHE effect. From symmetry considerations, in a system with broken-TRS, the BC satisfies $\mathbf{\Omega(\mathbf k)} = \mathbf{\Omega(-\mathbf k)}$, which leads to an intrinsic anomalous Hall effect (AHE). There also exist other types of materials where TRS is preserved, but $\mathcal{I}$ is broken, leading to $\mathbf{\Omega(\mathbf k)} =- \mathbf{\Omega(-\mathbf k)}$ with no AHE. Such symmetry configurations can be found in transition metal dichalcogenides~\cite{qian}, stacked graphene layers~\cite{mccann}, and semimetals~\cite{hasan}, where large BC values are observed near the band-crossing points. It is to be noted that MHE should be considered as non-linear Hall effect (rather than linear), because the non-linearity is manifested by the dependence
on the product of the in-built external field (i.e., the equilibrium field due to the source-drain potential difference), and an extra applied external bias while measuring the current \cite{papaj,agarwal}.

The central focus of this paper is to discuss the MHE in various three-dimensional (3D) semimetals which break TRS, in addition to the ones which preserve TRS.
Previously, MHE has been considered mainly in 2D platforms preserving $\mathcal T$ ~\cite{papaj,agarwal}. While 3D multi-Weyl systems have been considered very recently~\cite{das}, MHE was computed only for $\mathcal T$-preserving, $\mathcal I$-breaking materials. Moreover, there has not been much discussion about the effects of asymmetry (such as tilt or strain) on the different components of the MHE conductivity tensor.
In this paper, our aim is to provide a more complete study of 3D semimetals, considering situations covering all the above-mentioned aspects. As part of the new results, we consider a Weyl semimetal model which breaks TRS (cf. Sec.~\ref{sec_sweyl}), and then TRS-preserving multi-Weyl \cite{bernevig_mweyl}, multi-fold \cite{bradlyn,igor,ipsita-sajid}), and nodal-line \cite{Fang_2016} semimetals. A major aspect of our paper is the calculation of the MHE coefficients in presence of anisotropy (cf. multi-Weyl and nodal-line semimetals). We find that anisotropic dispersions, together with tilting of the nodes in various directions, can widely impact the Magnus Hall coefficients -- this is to the extent that the yield of one component is an order of magnitude larger than the other. This will be of great relevance for MHE experiments involving anisotropic media.

The paper is organized as follows. In Sec.~\ref{secderive}, we briefly review how the expression for MHE can be derived, using the Boltzmann transport formulation in the ballistic regime.
The analyses for the various semimetallic Hamiltonians are shown in Sec.~\ref{secresult}. In particular, we start by discussing the MHE for a TRS-broken Weyl semimetal in Sec.~\ref{sec_sweyl}. In Sec.~\ref{sec_mweyl}, we show how $\mathcal T$invariance and anisotropy affect the Magnus Hall transport. We also compute the features of MHE responses for multi-fold and nodal-line semimetals, in Sec.~\ref{secmultifold} and Sec.~\ref{secnodal}, respectively. In Sec.~\ref{secfinal}, we compare our results for the different semimetallic systems and discuss their implications. Finally, we end with a conclusion in Sec.~\ref{conclude}.

\section{Review of the Derivation of the MHE}
\label{secderive}

\begin{figure}[]
	\centering
	\includegraphics[width =0.46 \columnwidth]{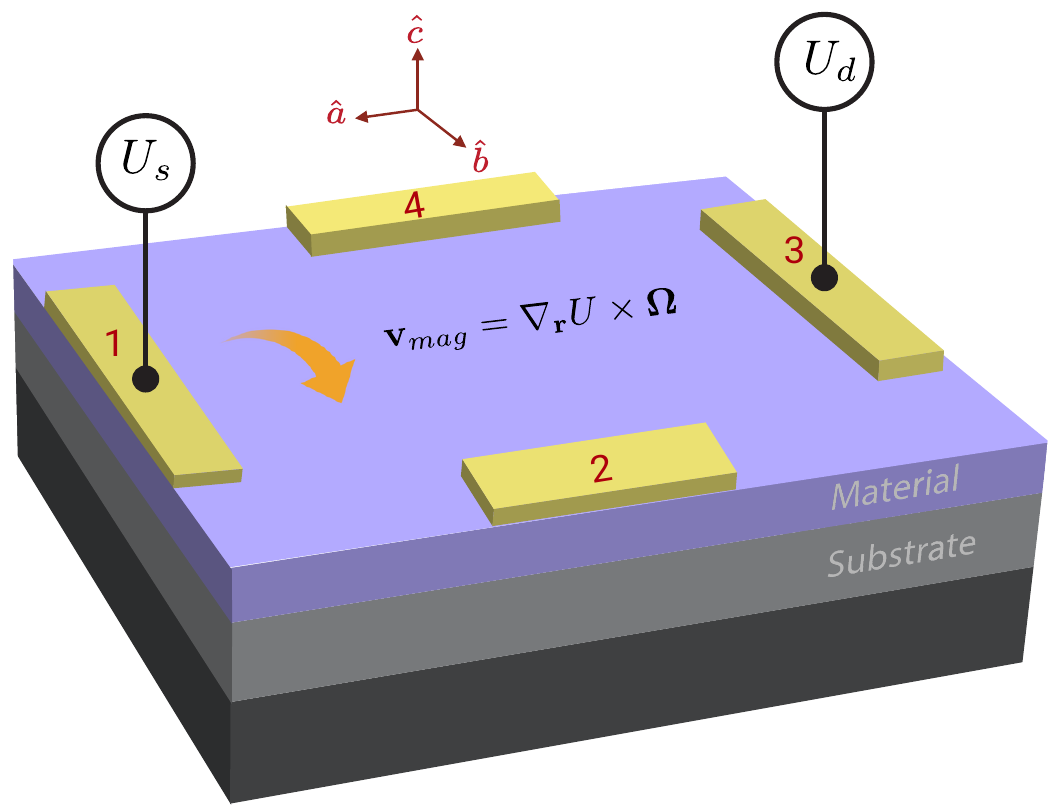}
	\caption{
\label{fig:cartoon}	
The figure shows a schematic representation of the Magnus Hall set-up. Due to the potential difference $\Delta U= U_d-U_s$ across the length of a Hall sample, an in-built electric field
${\mathbf E}_{in}= e^{-1}\partial_a U \, \hat{a}$ exists between the source ($\mathbf 1$) and the drain ($\mathbf 3$). This electric field couples to the BC-component $ \Omega_c $ (along $\hat c$) in a way such that when an electron exits the source, it gets deflected towards $\hat{b}$, generating a current along that direction. This transverse current can be measured by attaching a probe between the contacts $\mathbf 2$ and $\mathbf 4 $.}
\end{figure}

We consider a 3D mesoscopic slab, with the orthogonal axes labelled as $a$, $b$, and $c$. In particular, the slab is assumed to be extended infinitely in the $\hat{a}$-direction (the hat denotes a unit vector in the corresponding direction), but has a finite width along $\hat{b}$. We then apply a slowly-varying potential energy across its length, i.e., along $\hat{a}$. This introduces a bias voltage $\Delta U$ between two ends of the sample, namely the source and the drain, which creates an in-built electric field $\*E_{in}=(\frac{1}{e}\, \partial_a U, \, 0, \, 0)$. Additionally, there is an applied electric field $\boldsymbol{\mathcal{E}}_{app}$ that drives the electrical current. The motion of the electrons in this sample is dictated by the following two coupled equations:
\begin{align} 
	\label{eq:rdot}                                                                                                       
\hbar \, \dot{{\mathbf r}} =                                                                                                  
\mathbf v_{band} \, +  \left (\nabla_{\mathbf r} U
+  e \,\boldsymbol{\mathcal{E}}_{app} \right ) 
\times \mathbf \Omega \,, \quad 
	\hbar \, \dot{{\mathbf k}} = -\nabla_{\mathbf r} U -\, e \, \boldsymbol{\mathcal{E}}_{app}\,,                                         
\end{align}
where $\mathbf v_{band}= \nabla_{{\mathbf k}} \mathcal E$ is the band velocity, $\mathcal E({\mathbf k})$ is energy dispersion, and $\mathbf{\Omega}$ is the BC. The terms $ \nabla_{\mathbf r} U \times \mathbf  \Omega$ and $\boldsymbol{ \mathcal{E} }_{app}\times \*\Omega$ define the Magnus velocity $\* v_{mag}$ and the anomalous velocity $\mathbf v_{ano}$, respectively. The former ($\*v_{mag}$) arises from a non-zero value of $\partial_a U$, as electrons that are moving from source to drain will get drifted towards $\hat{b}$ with the value $\Omega_c \, \partial_a U  $. This generates a Hall current proportional to the $c$-component of the BC, which can be estimated by measuring the transverse voltage (cf. Fig.~\ref{fig:cartoon}). We aim to understand the Hall conductivity for ballistic transport regime, when  the scattering time $\tau$ is so large that effectively no collision takes place within the sample. 

To calculate the MHE conductivity tensors, we use the Boltzmann transport equation~\cite{lundstrom, dresselhaus}, that describes the evolution of the distribution function $f(\mathbf{k}, \mathbf{r},t)$ of the carriers as follows:
\begin{align} 
	\label{eq:BTE}                                                                                                                                                    
	\left( \frac{\partial}{\partial t} + \, \dot{{\mathbf r}}\, . \nabla_{{\mathbf r}} + \, \dot{{\mathbf k}} \, . \nabla_{{\mathbf k}} \right) f({\mathbf k}, {\mathbf r},t) = I_{coll} [ {f({\mathbf k}, {\mathbf r} ,t)} ]\,. 
\end{align}
The term on the right-hand side arises due to electron scatterings, and thus can be set to zero in the ballistic regime. We also restrict ourselves to the steady-state, so that $\partial f/\partial t = 0$.
The carrier distribution function can be decomposed as $f(\mathbf{k}, \mathbf{r},t)= f_0 + f_1$, where $f_0$ is the equilibrium value given by the Fermi-Dirac distribution function $f_0 =  \frac{1}
{1+ \, e^{\frac{ \tilde {\mathcal E}({\mathbf k}, {\mathbf r}) - \bar{\mu} } { k_B \, T} }}$, and $f_1$ is the deviation from $f_0$ (i.e., it refers to the non-equilibrium part).
Here, $ \tilde {\mathcal E} ({\mathbf k}, {\mathbf r}) = \mathcal E({\mathbf k}) + \, U({\mathbf r})$, and $\bar{\mu}$ is the fixed (equilibrium) electrochemical potential. For the expression of $f_1$, we take an ansatz~\cite{papaj}
\begin{align}
	f_1({\mathbf k},{\mathbf r})=
	\begin{cases}
	-\Delta \mu \, \partial_{\mathcal E} f_0 
	- \frac{\tilde {\mathcal E}({\mathbf k}, {\mathbf r})-\mu}{T} \, \Delta T \, \partial_{\mathcal E} f_0 & \text{ for } v_a>0      \\ 
	0                                                                                      & \text{ for } v_a \leq 0 
	\end{cases}\,,	
\end{align}
where $\mu$ is the total chemical potential.
The ansatz is motivated through the fact that the bias voltage introduces an imbalance $\Delta \mu $ in the electrochemical potential, between the source and the drain. Due to this imbalance, only electrons with $v_a>0$ flows from source to drain, creating a steady current in the process. In addition, a thermal gradient $\Delta T $ may exist between the two ends, leading to a thermal conduction as well. 
Equipped with this, we can finally compute the electric ($\*J$) and thermal ($\*Q$) currents, using the expression
\begin{align}
	\{\*J, \, \*Q \} = \int \, \frac{d^3k}{(2\pi)^3} \, \dot{{\mathbf r}}\, \{-e, \, 
	\tilde {\mathcal E} ({\mathbf k},{\mathbf r})-\mu \} \, f_1({\mathbf k},{\mathbf r},t) \,.               
\end{align}

Let us now consider the Hall set-up oriented in such a way that the Magnus Hall current is in the
negative $ \hat b$-direction, as described in the first paragraph above. For TRS-invariant materials, the anomalous contributions in Eq.~\eqref{eq:rdot} cancel out, and the $a$-component of $\dot{{\mathbf r}}$ gets contribution from the band-velocity and the Magnus term only. With this prescription, we get
\begin{align}
	J_b & = -e \int \, \frac{d^3k} {(2\pi)^3} 
	\left( \dot{\mathbf r} \right)_b \, f_1({\mathbf k},{\mathbf r}) 
	=	\left[ \mathcal{J}_{b} \, + \, \mathcal{J}_{b}^{'} \right]_{elec} + \left[ \mathcal{J}_{b} \, + \, \mathcal{J}_{b}^{'} \right]_{th} \,, 
\end{align}
giving rise to electric (superscript ``$elec$'') and thermal (superscript ``$th$'') currents. Furthermore, each of these currents have two types of contributions -- one from the Magnus velocity (primed), and another from the band-velocity (unprimed). Since the current density $J_b$ has a spatial dependence, an integration over the length between the gates ($L$) is required to obtain the current $I_{b}$. The expession $I_{b} = \frac{1}{L} \, \int_{0}^{L} J_b$ is then used to get the average conductivity, defined as $\sigma_{ab} = \frac{e\, I_{b}}{\Delta \mu}$. Let us denote the electrical, Nernst, and thermal conductivity tensors by the symbols $\sigma_{ab}$, $\alpha_{ab}$, and $\kappa_{ab}$, respectively.
We use the superscript ``$0$'' when these conductivities are associated with the semi-classical band-velocity parts, such that
\begin{align} 
\label{eq:unprimed}
\left \lbrace \sigma_{ab}^0 \, , 
\,\, \alpha_{ab}^0 \, , \,\, \kappa_{ab}^0
\right \rbrace 
= -L\, \int \frac{d^3k}{(2\pi)^3} \,  
\left \lbrace
e^2\,, \,\, -\frac{e\left (\mathcal E-\mu \right )}{T} \,,\,\,
\frac{\left( \mathcal E-\mu \right )^2}{T}  \right \rbrace  
\partial_{\mathcal E} f_0 \,,
\end{align}
On the other hand, the Magnus Hall conductivities take the form
\begin{align} 
\label{eq:primed}
\sigma_{ab}^{m} & = - \frac{e^2\,  \Delta U \, \epsilon_{abc} } {\hbar}
 \int_{v_a>0} \frac{d^3k} {(2\pi)^3} \, \Omega_c \, \partial_{\mathcal E} f_0                        
 \,,\nonumber \\
\alpha_{ab}^{m}  & = \frac{e \, \Delta U \, \epsilon_{abc} } {\hbar\,T} 
 \int_{v_a>0} \frac{d^3k} {(2\pi)^3} \, \Omega_c \, (\mathcal E \, - \, \mu) \, \partial_{\mathcal E} f_0 \,,\nonumber \\
\kappa_{ab}^{m} & = - \frac{\, \Delta U \, \epsilon_{abc}} {\hbar\, T}  
\int_{v_a>0} \frac{d^3k} {(2\pi)^3} \, \Omega_c \, (\mathcal E \, - \, \mu)^2 \, \partial_{\mathcal E} f_0 \,,
\end{align}
where we have used the superscript ``$m$'' in order to denote their association with the MHE.
Here, $\Delta U = \int_0^L \, da \, \partial_a U $ is the average potential difference along $\hat{a}$, and $\epsilon_{abc}$ denotes cyclic permutation (i.e., the Levi-Civita symbol). An important difference to note is that while the transport coefficients in Eq.~\eqref{eq:unprimed} are independent of $\Delta U$, the Magnus Hall conductivities in Eq.~\eqref{eq:primed} vary linearly with $\Delta U$ -- this is the reason why MHE is a non-linear effect. Due to the factor $\partial_{\mathcal E} f_0$ in each integrand, states near the Fermi surface (FS) take part in conduction. More explicitly, the conduction depends on the carriers with $v_a({{\mathbf k}})>0$ in the vicinity of the FS, and how they activate the BC. If the Fermi energy is much higher than the thermal energy, then $\sigma_{ab}^{m}$, $\alpha_{ab}^{m}$, and $\kappa_{ab}^{m}$ are related to each other by the Wiedemann-Franz law and the Mott relation, as shown below:
\footnote{The Wiedemann-Franz law states that the ratio of the thermal and electrical conductivities is equal to $ {\mathcal L}\,T$ (where ${\mathcal L}$ is the Lorentz number), and this statement holds only if the Fermi energy (i.e., the chemical potential $\mu$) is much larger than the thermal energy ($\sim k_B \,T$).
In other words, the Wiedemann-Franz law is not expected to hold unless $\mu \gg k_B\, T$  (with $k_B T \, \sim 20$ meV). This explains the disparity between $\sigma^m$ and $\kappa^m$ as $\mu \rightarrow 0$, which is visible in our results for all the semimetals considered here.}
\begin{align} 
\label{eq:franz-mott}                                                       
\alpha_{ab}^{m} = - \frac{\pi^2 \,k_B^2 \,T} {3\, e} \,                       
	\frac{\partial \sigma_{ab}^{m}}{\partial \mu} \,,\quad                      
	\kappa_{ab}^{m} = \frac{\pi^2\, k_B^2\, T} {3 \,e^2} \, \sigma_{ab}^{m} \,. 
\end{align}
An important feature is that each transport coefficient is an odd function of $\mu$. This means that either lowering or raising the Fermi level is sufficient to obtain the same magnitude of the response, albeit with opposite signs.
For this reason, we only consider the chemical potential cutting the (1) valence bands for the nodal point semimetals, and (2) conduction band for the nodal-line semimetal.

\section{Magnus Hall Response for Various 3D Semimetals} 
\label{secresult}

In this section, we present the numerical results for the Magnus Hall transport, considering different types of 3D semimetals. We have used \textit{Mathematica} to compute the Magnus Hall conductivities. In particular, we have used the spherical polar coordinates (except for the nodal-line case), and the `GlobalAdaptive' strategy, to compute the integrals as a function of the chemical potential $\mu$ (with a scanning interval of $\mathrm{0.05}$ eV). For such numerical calculations, it is always important to make sure that the results do not depend on the unphysical parameters, such as the integration cut-off. Ideally, the cut-off is experimentally motivated. But since we focus here on a more generic description of the semimetals (based on low-energy continuum models), instead of specific materials, we have chosen the FS cut-off through a trial and error process by checking the convergence of the integrals. We have also used the natural units, by setting $\hbar =c =e =1$ throughout, for the sake of simplicity.

\subsection{Weyl Semimetal} 
\label{sec_sweyl}

\begin{figure*}[] 
	\centering
	\includegraphics[width=\columnwidth]{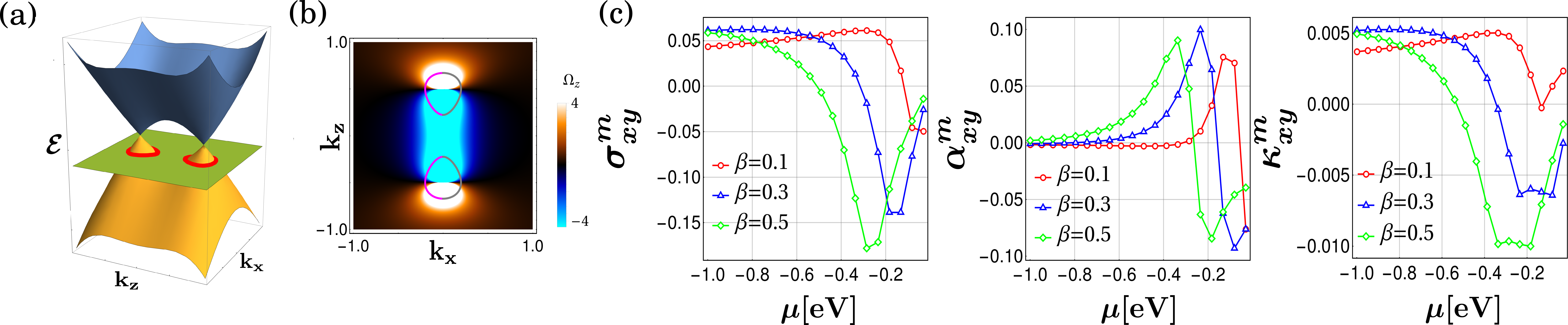}
	\caption{\textbf{(a)} Energy dispersion of the Weyl Hamiltonian in Eq.~\eqref{eq:TbrokenWeyl} for $\mu =-0.2$ eV and $\beta=0.5$ {eV}$^{-1}$, with the projection of the 3D FS highlighted in red. \textbf{(b)} The $z$-component of the BC is shown, along with the superimposed projections of the FS contours in the $k_z$-$k_x$ plane. The latter is color-coded such that the states with $v_x>0$ are in magenta, and those with $v_x<0$ are in gray. \textbf{(c)} The Magnus Hall coefficients
$\sigma^m_{xy}$ (in units of $10^{-3}$ eV), $\alpha^m_{xy}$ (in units of $10^{-3}$ eV$^2$/K), and $\kappa^m_{xy}$ (in units of $10^{-3}$ eV$^3$/K) are displayed as functions of $\mu$, for three different values of $\beta$. Here, we have set $\Delta U = 0.05$ eV and $T=300$ K.}
	\label{fig:weyl}
\end{figure*}

Weyl semimetallic phases appear when the $\mathcal{T}$ or $\mathcal{I}$ is broken in a material. As a first step, $\mathcal{T}$-broken semimetals are easiest to think of, since they hold the minimum number of Weyl nodes, resulting in a total of two nodes of opposite chiralities. In order to discuss the MHE in the absence of $\mathcal{T}$, we consider a $2\times 2$ Weyl Hamiltonian given by~\cite{weyl-rev,flores}
\begin{align} \label{eq:TbrokenWeyl}
\mathcal{H} = v_F \, \boldsymbol{\sigma} \cdot \mathbf{d_k} \,, 
\end{align}
where $\boldsymbol{\sigma}$ is the vector of the three Pauli matrices, $\mathbf{d_k} = \{ k_x, \, k_y, \,\frac{ k_z^2-\beta^2 } {2\,\beta} \}$, and $v_F$ (set to unity hereafter) is the Fermi velocity. The Hamiltonian parameter $\beta$ separates the Weyl nodes in the momentum space, so that the band-crossing points occur at $(0,0, \pm \beta)$ [cf. Fig.~\ref{fig:weyl}(a)]. In particular, the two Weyl points are separated by a distance of $2 \beta$. Since an unbounded linear dispersion is not realistic in solid state systems, we have used the model from Ref.~\cite{flores}, where band-bending effects have been introduced by taking the dispersion along the $z$-direction to be quadratic.

The broken-$\mathcal{T}$ is easy to envision because of the node-splitting~\cite{burkov}, as the Weyl points at $k_z=\pm \beta$ are of opposite chiralities, and do not obey the Kramer's theorem. The FS can be obtained by solving $\mathcal E({\mathbf k})= \mu$, where $\mu$ is the chemical potential, $\mathcal E({\mathbf k}) \equiv \pm \, \epsilon_{{\mathbf k}} $ is the energy dispersion, $\epsilon_{{\mathbf k}} =  \sqrt{k_{\perp}^2 + k_{||}^2}$, $k_{\perp} = \sqrt{k_x^2+k_y^2}$, and $k_{||} = 
\frac{ k_z^2-\beta^2 } {2\,\beta}$.
As we move to low energies (${\mathbf k} \rightarrow 0$), higher-order terms become negligible, and the dispersion appears to be linear in all directions to the first approximation. Here we show the behaviour of the $xy$-component of the MHE conductivities, and hence the $x$-component of band-velocity ($v_x$) and the $z$-component of BC ($\Omega_z$) are relevant. These can be expressed as:
\begin{align}	
	\label{eq:BC}
	v_x ({\mathbf k}) & = \partial_{k_{x}} \mathcal{E}({\mathbf k})= \pm \, \frac{k_x}{\epsilon_{\mathbf k}}                                                                                                                                                      \,, \quad 
\Omega_z ({\mathbf k})  = \pm \, \frac{1}{2} \, \*{\hat{d}_k} \cdot
 \left (\partial_{k_x} \, \*{\hat{d}_k} \times \partial_{k_y} \, \*{\hat{d}_k} \right )
 = \pm \, \frac{k_{||}}{2 \, \epsilon_{\mathbf k}^3}, \quad 
 {\mathbf{\hat{d}}_{k}} = \frac{\*{d_k}}{|\*{d_k}|} \,.
\end{align}
In the Fig.~\ref{fig:weyl}(a), we show the band-structure for $\beta=0.5$ {eV}$^{-1}$, with the projection of the FS shown in red color for $\mu=-0.2$ eV. As mentioned earlier, the interplay of the FS and BC plays a central role in deciding the fate of the MHE response. This can be understood through Fig.~\ref{fig:weyl}(b), which shows the FS contours superimposed on the contour-plot for $\Omega_z$ in the $zx$-plane. The FS is color-coded according to its slope, such that the contributing states with $v_x>0$ are in magenta, and the inert states with $v_x<0$ are colored gray. When $\mu$ is tuned to cut the Weyl nodes, the FS structure is point-like, existing at the two points $k_z = \pm \beta$. Upon lowering $\mu$, the FS transforms into two separate closed manifolds in the BZ, with the Chern numbers $\pm 1$. For $\mu > \pm \beta/2$, the FS undergoes a topological Lifshitz transition, in the sense that two separate lobes merge into a single surface. It is useful to emphasize that any MHE response depends on the states near the FS, due to which a change in its topology dramatically affects the conductivities. We capture this in Fig.~\ref{fig:weyl}(c), where we plot $\sigma^m_{xy}$, $\alpha^m_{xy}$, and $\kappa^m_{xy}$ as functions of $\mu$, for $\beta=0.1, 0.3, 0.5$. We observe that for $\beta=0.5$, $\sigma^m_{xy}$ starts to increase, attaining a peak at the Lifshitz transition, and then falling off as the Fermi level is brought closer to zero. This is because, away from the Weyl nodes, the FS mostly activates vanishing BC, which leads to a small response. As $\mu$ is raised and brought closer to zero energy, although the FS shrinks, it activates larger values of BC, which in turn increases the response. Finally, the response decreases when $\mu$ is close to the nodes, which can be attributed to the cancellation of the contributions coming from Fermi pockets in that regime.  

\subsection{Multi-Weyl Semimetals} 
\label{sec_mweyl}

Ideally for Weyl-like excitations, the band-dispersion is linear in every direction in the low-energy limit. But this can be too strong a constraint in real materials, as anisotropy is quite ubiquitous. One particular scenario is when the dispersion is either quadratic or cubic in the $xy$-plane, and linear along the $z$-direction, which we usually refer to as ``multi-Weyl'' semimetals~\cite{bernevig_mweyl}. These systems contain a higher value of the monopole charge (i.e., the total Chern number of either the valence or the conduction bands) than that of the isotropic Weyl semimetal. For example, a monopole charge of $J =2$ (double-Weyl nodes) can be seen in  $\mrm{HgCr_2Se_4}$~\cite{zhong} and $\mrm{SrSi_2}$~\cite{huang}, whereas materials with composition $\mrm{A(MoX)_3}$ (with A=Rb, Tl, and X=Te)~\cite{zunger} harbour $ J =3$ monopole charges. They can also be found as  Bogoliubov-de Gennes (BdG) quasiparticles in the superconducting states of materials like $\mrm{UPt_3}$~\cite{andriy}, $\mrm{SrPtAs}$~\cite{neupert}, and $\mrm{YPtBi}$~\cite{roy}. We consider the generic low-energy Hamiltonian~\cite{trescher}
\begin{align} \label{eq:mweylHam}
&\mathcal{H}=                                       
	\sum_{a} v_{aa}^{\prime}  \, k_a    
	+ \sum_{a,b} \, v_{ab} \, d_a({\mathbf k}) \, \sigma_b \,, 
\quad d_x= k_{\perp}^J \cos(J\phi), \, d_y = k_{\perp}^J \sin(J\phi), \, d_z =  k_z\,,	\nonumber \\
& k_{\perp} = \sqrt{k_x^2+k_y^2}\,,\quad
\phi = \arctan \left (k_y/k_x \right ),
\end{align}
which captures such anisotropy.
Here, $a,b = \{ x, y, z\}$ spans over the three Cartesian axes, and $J$ is an integer that defines the order of the dispersion in the $xy$-plane -- $J=2$ and $J=3$ denote quadratic and cubic dispersions, respectively. The first term in Eq.~\eqref{eq:mweylHam} signifies the tilting of the cone. For simplicity, we choose $v_{ab}= v_{F} \, \delta_{ab}$, and set $v_{F}=1$. For $v_{aa}^{\prime}= \zeta \, \delta_{az}$, the energy eigenvalues are given by $\mathcal E ({\mathbf k}) = \zeta \, k_z \pm \, \epsilon_{{\mathbf k}}$, with
$\epsilon_{{\mathbf k}}= \sqrt{k_{\perp}^{2J}+ k_z^2} $.

In our numerics, we consider a single cone with positive chirality, as the calculation is similar for a node with negative chirality. It is useful to note that since $d_a({{\mathbf k}})$ directly affects the eigenspinors, any anisotropy in $d_a({{\mathbf k}})$ is reflected in the BC. This implies that the values of the MHE transport coefficients will be different depending on which directions we are probing. With this in mind, here we illustrate both the $xy$- and $zx$-components of the MHE conductivities.

\begin{figure}[] 
	\centering
	\includegraphics[width=\columnwidth]{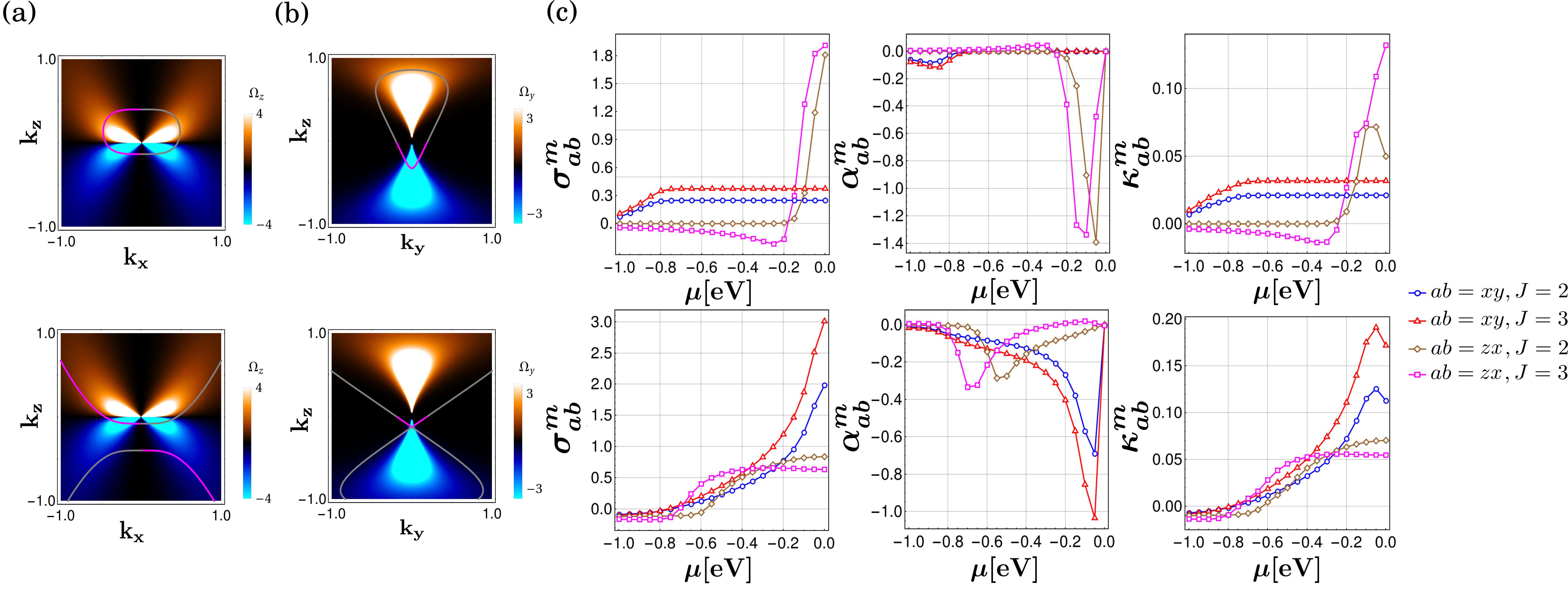}
	\caption{\label{fig:mweyl}
For a single node of a multi-Weyl semimetal with $J=3$, the left-hand side panels show the FS projections ($\mu=-0.2$ eV), superimposed on the components of the BC, relevant for computing \textbf{(a)} the $xy$-component, and \textbf{(b)} the $zx$-component of the Magnus Hall transport coefficients. The upper panels of \textbf{(a)} and \textbf{(b)} are for the type-I phase with $\zeta=0.5$, and the lower panels are for the type-II phase with $\zeta=1.5$. The magenta-colored regions of the FS projections refer to the states with $v_a >0$, while the gray-colored areas indicates $v_a <0$, where $a = x$ for subfigure \textbf{(a)}, and $a =z$ for subfigure \textbf{(b)}. 
The curves in subfigure \textbf{(c)} show different components of the Magnus Hall coefficients $\sigma_{ab}^m$ (in units of $10^{-3}$ eV), $\alpha^m_{xy}$ (in units of $10^{-3}$ eV$^2$/K), and $\kappa^m_{xy}$ (in units of $10^{-3}$ eV$^3$/K), as functions of $\mu$, for $\zeta=0.5$ (upper panel) and $\zeta= 1.5$ (lower panel). Here, we have set $\Delta U=0.05$ eV and $T=300$~K.}
\end{figure}

The MHE response in different directions can be either zero or finite, depending on the direction of tilt. This becomes apparent from the explicit expression of the BC, given by
\begin{align}
\mathbf	\Omega({\mathbf k})  = \pm \frac{J \, k_{\perp}^{2(J-1)}}{2 \, \epsilon_{\mathbf k}^3} 
\{k_x, \, k_y, J\, k_z \}   \,.         
\end{align}
Because of unbroken $\mathcal T$ and broken-$\mathcal{I}$, the components of the BC satisfy
$\Omega_a \propto k_a$. 
Therefore, if we want to probe $\Omega_z$, the $k_x$-$k_z$ and $k_y$-$k_z$ projections of the FS are relevant, as $\Omega_z$ is non-zero in those planes. Consequently, a tilt at least along the $k_z$-direction is needed to make the FS asymmetric, so that the MHE response does not get cancelled over the BZ. This tilt can be implemented by setting $v_{aa}^{\prime}= \zeta \, \delta_{az}$, and the band velocity is then given by $\mathbf	v({\mathbf k})  = \{ \pm \frac{J \, k_{\perp}^{2\left (J-1 \right )}  \, k_x}
{\epsilon_{\mathbf k}}, 
\pm \frac{J \, k_{\perp}^{2\left (J-1 \right )} \, k_y} {\epsilon_{\mathbf k} }, \, 
\zeta \pm \frac{k_z}{\epsilon_{\mathbf k}} \} \, \quad $
Similarly, we set $v_{aa}^{\prime}=\zeta \, \delta_{ay}$ for the scenario when $\Omega_y$ is probed. Note that although the computation of the $xy$-components of the conductivity tensors have been demonstrated recently in Ref.~\cite{das}, we have included it here for the sake of completeness. 

Although conductivity depends on a 3D integral, the fate of the individual components solely relies on a single FS projection. For instance, for the case of the $xy$-component, the relevant FS projections are in the $k_x$-$k_z$- and $k_y$-$k_z$ planes, since $\Omega_z$ is probed. Out of these two, only $k_x$-$k_z$ projection is important as $v_x$ is zero in the $k_y$-$k_z$-plane. Similarly, for the $zx$-component of conductivity, only FS projection in the $k_y$-$k_z$-plane is important. These two relevant planes are shown in Fig.~\ref{fig:mweyl}(a) and (b) for $J=3$. 
We present our MHE results in Fig.~\ref{fig:mweyl}(c) for both the $xy$- and the $zx$-components. In the upper panel of Fig.~\ref{fig:mweyl}(c), when the Fermi level is close to the node and the tilt is sub-critical, the $xy$-component of the Hall response is nearly constant but the $zx$-component rises dramatically. This contrasting behaviour can be traced back from the Fig.~\ref{fig:mweyl}(a) and \ref{fig:mweyl}(b), which show stark differences in the relevant FS projections for the two cases. We also notice that the response is always larger for cubic dispersion, as BC strength is proportional to $J$. Increasing the amount of tilt from $\zeta=0.5$ to $\zeta=1.5$ over-tilts the cone, and makes the FS open. This affects the MHE response because the $xy$-component, which was constant earlier, now steadily rises as $\mu \rightarrow 0$. But the $zx$-component remains small owing to the small active FS.

\subsection{Multi-fold Semimetals} 

\label{secmultifold}

Going beyond the two-band semimetals discussed so far, we now focus on multiple band-touching points in the BZ \cite{bradlyn}, which we generically call multi-fold semimetals. The band degeneracy point of such semimetals is predicted to host excitations of higher-pseudospin quasiparticles, some of which have no analogues in high-energy physics. So far excitations with pseudospin values of $s=1$ and $s=3/2$ have been predicted / observed. The latter is also known as Rarita-Schwinger-Weyl (RSW) semimetals \cite{igor,ipsita-sajid}. Experimental observations include signatures of three-fold degeneracy point at the center of the BZ in $\mathrm{CoSi}$~\cite{takane}, and a combination of three-fold and four-fold degeneracies at the $\Gamma$ and $R$ points of $\mathrm{RhSi}$~\cite{sanchez}. The linearized $\mathbf k \cdot \mathbf p$ Hamiltonian for such systems can be written as
\begin{align}
	\mathcal{H} =   \zeta \, k_z +   \mathbf{k} \cdot \mathbf{S} \,, 
\end{align}
where $\mathbf{S}$ refers to the vector comprising the three matrices representing pseudospin $s$. The first term represents a tilt along the $z$-direction.

\begin{figure*}[] 
	\centering
	\includegraphics[width=\columnwidth]{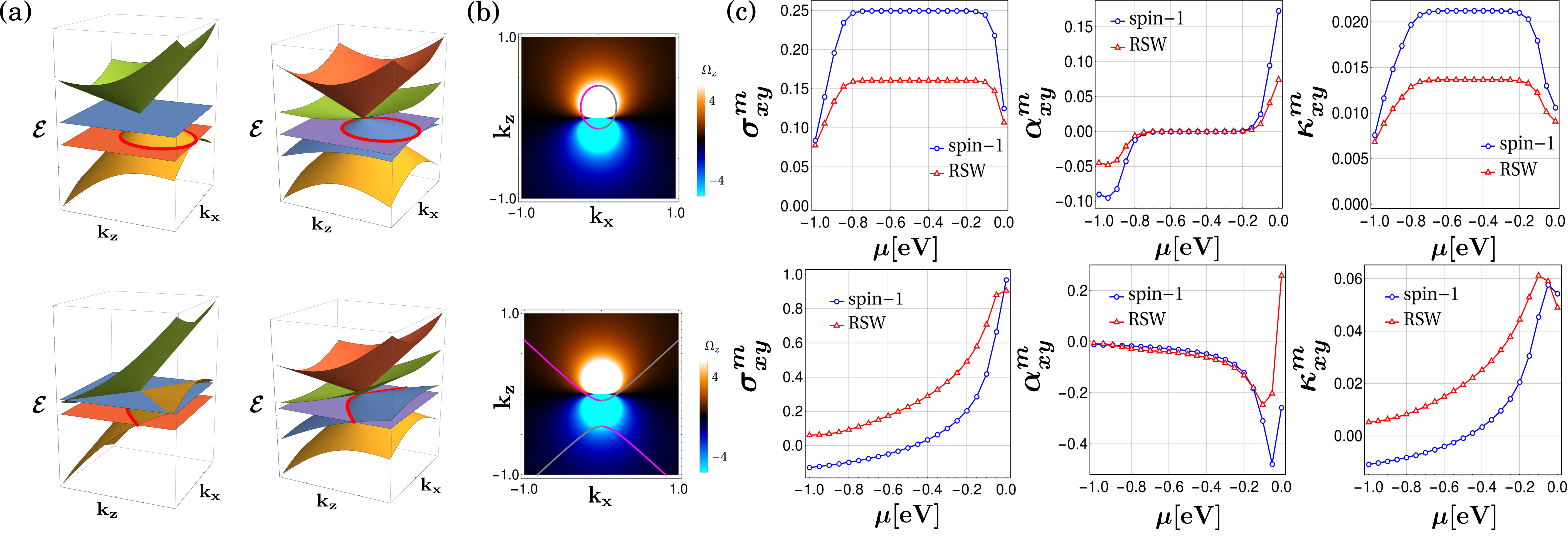}
	\caption{\label{fig:multifold}
\textbf{(a)} Dispersions of the pseudospin-1 and RSW semimetals are displayed side by side for $\mu = -0.2$ eV (with $\zeta=0.5$ and $\zeta =1.5$ for the upper and lower panels, respectively). \textbf{(b)} FS contours superimposed on the $z$-components of the BC are shown for the pseudospin-1 quasiparticles. The FS projections have the same color-coding as in the earlier figures. \textbf{(c)} The plots show the MHE coefficients $\sigma^m_{xy}$ (in units of $10^{-3}$ eV), $\alpha^m_{xy}$ (in units of $10^{-3}$ eV$^2$/K), and $\kappa^m_{xy}$ (in units of $10^{-3}$ eV$^3$/K), as functions of $\mu$. The upper panel is for the type-I phase with $ \zeta =0.5$, and the lower panel is for the type-II phase with $ \zeta=1.5$. Here, we have used $\Delta U = 0.05$ eV and $T=300$ K.}
\end{figure*}

For $s=1$, the energy eigenvalues of $\mathcal E({\mathbf k})=0, \pm k$ include three bands. Out of these three, two are topological with Chern numbers $ {C}=\pm 1$, and the third is a trivial flat-band with $ {C}=0$. Owing to the flat-band, which does not take part in transport, the transport mechanism of pseudospin-1 quasiparticles is essentially similar to a Weyl semimetal (which has two bands). On the other hand, the RSW case with $s=3/2$ has four topological bands, with $ {C}=\pm 1, \pm 3$, resulting in a high monopole charge of magnitude $ 4$. For computing the $xy$-components of the conductivity tensors, we need the following components of the band-velocity and BC:
\begin{align}
	v_x^p       = p \, \frac{k_x}{2\,k} \,,\quad  
	\Omega_z^p  = \pm p \, \frac{k_z}{2\,k^3} \,, 
\end{align}
respectively. Here, $p$ takes the value $2$ for the pseudospin-1 system, and the values $\lbrace 1,3 \rbrace $ for the RSW Hamiltonian, and we have used the subscript ``$p$'' to indicate which case we are considering.

The band-structures, FS projections, $z$-components of the BC, and the MHE responses are shown in Fig.~\ref{fig:multifold}. Naively one would expect the MHE conductivity tensors to be larger in magnitude due to the presence of more bands. However, our results in Fig.~\ref{fig:multifold}(c) indicate that this is not always true. For example, even though RSW semimetal has more bands than the pseudospin-1 model, only two out of the four bands essentially participate in conduction, when tilting is small (i.e., in the type I phase). 
Moreover, for RSW quasiparticles, the bands responsible for conduction have a BC value that is half of the BC value of the corresponding pseudospin-1 bands. As a result, in the type-I phase, the MHE response of the pseudospin-1 quasiparticles is larger than that of the RSW quasiparticles for a representative value of $ \zeta =0.5$. However, tuning to the type-II phase changes the scenario drastically. Over-tilting the cones makes all the bands have non-zero FS projections. Therefore, in this regime, the MHE conductivity tensor components of RSW quasiparticles are larger in magnitude than those of the pseudospin-1 quasiparticles.

\subsection{Nodal-line Semimetals} 
\label{secnodal}

\begin{figure*}[] 
	\centering
	\includegraphics[width=\columnwidth]{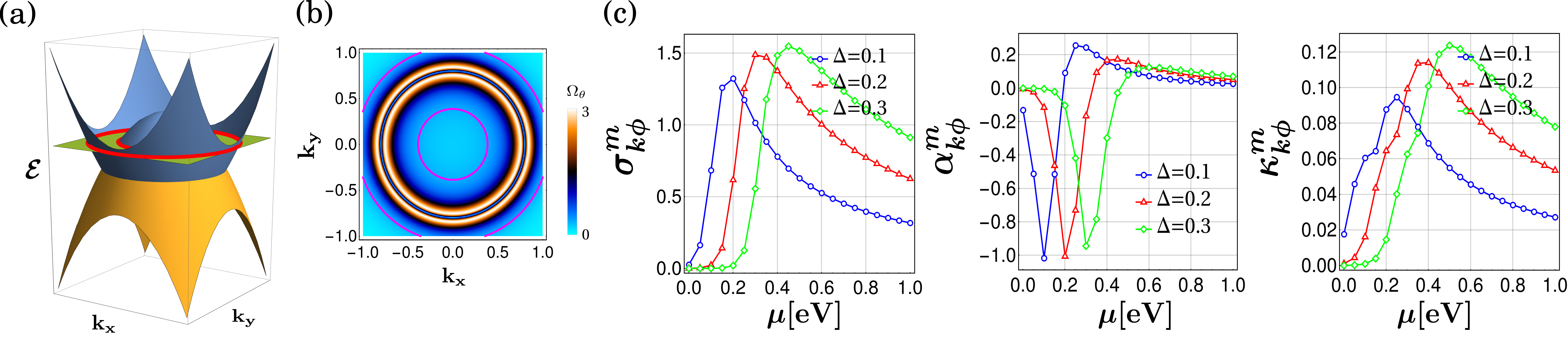}
	\caption{\label{fig:nodal}
Subfigure \textbf{(a)} shows the dispersion of a nodal-line semimetal, given by Eq.~\eqref{hamnode}, for $ \mu = 0.3$ eV. The Fermi level is shown in red.
\textbf{(b)} The density plot of the $\theta$-component of the BC is shown, along with the projections of the FS, for the conduction band. The magenta color refers to the Fermi states with $v_k >0$. \textbf{(c)} The plots show the MHE coefficients $\sigma_{k\phi}^m$ (in units of $10^{-3}$ eV$^2$), $\alpha_{k\phi}^m$ (in units of $10^{-3}$ eV$^3$/K), and $\kappa_{k\phi}^m$ (in units of $10^{-3}$ eV$^4$/K), as functions of $\mu$, for three different values of the SOC strength $\Delta$. Here, we have used the parameter values $\Delta U=0.05$ eV, $T=300$ K, $v=2$, $k_0=0.8$ {eV}, and $\lambda=1$ {eV}$^{-1}$.}
\end{figure*}

In addition to semimetals with bands crossing at a point, other classes of semimetals are possible where band-touching occurs along nodal-lines \cite{Fang_2016}. These are known as nodal-line semimetals (NLSMs). The nodal ring of such materials can be protected by chiral symmetry
$\mathcal{C}$, mirror symmetry $\mathcal{M}$, or a combination of particle-hole and time-reversal symmetries (denoted by $\mathcal{PT}$). However, all of these symmetries force the BC to vanish~\cite{liu}, and a mass term is typically needed to make the BC non-trivial. This mass term can appear in the form of weak SOC. In presence of the SOC, an extra glide symmetry is needed, along with $\mathcal{P T}$, to protect the nodal lines (cf. in $\mathrm{SrIrO}_3$~\cite{chen,ashvin}). For our purpose, we consider a minimal two-band Hamiltonian~\cite{yang} with SOC, that has a single nodal-line as follows:
\begin{align}
	\label{hamnode}                                                            
	\mathcal{H} =   \lambda  \left( k_{\perp}^2 \, - \, k_0^2 \right) \sigma_x 
	+   v \, k_z \, \sigma_y + \Delta \, \sigma_z \,,                          
\end{align}
where the first term depicts a circular node $k_x^2 \, + \, k_y^2 = k_0^2$ in the $k_x$-$k_y$ plane, and $\Delta$ is the strength of the SOC.
Unlike the $\mathcal{T}$-invariant cases considered earlier, here we do not require a finite tilt to observe a non-zero MHE response. This is because, owing to the underlying symmetry of the Hamiltonian, the relevant BC component is either positive or negative over the entire BZ, and there is no scope for internal cancellations [see Fig.~\ref{fig:nodal}(b)]. For $\Delta=0$, the Hamiltonian exhibits two discrete topological symmetries: (1) $\mathcal{C} $, which can be represented by $\sigma_z$, and which acts as $\mathcal{C\,H\,C}^{-1}=-\mathcal{H}$;
(2) $\mathcal{PT} $, which can be represented by $ \sigma_x \, \mathcal{K}$ (where $\mathcal{K}$ is complex conjugation), and which acts as $\mathcal{(PT)\, H\, (PT)}^{-1}=\mathcal{H}$. Since our interest is in the low-energy states, the above Hamiltonian can be linearized by a transformation into the toroidal coordinates as follows:
\begin{align}
	k_x & = \left (k_0 \, + \, k \, \cos\phi \right)  \cos\theta \,, \quad 
	k_y  = \left (k_0 \,+ \,k \, \cos\phi \right )  \sin\theta \,,\quad
	k_z  = k \, \frac{\sin\phi}{\alpha}  \,,            
\end{align}
where $k$ is radius of the toroid, $k_0$ is the radius of the toroid from its center, $\alpha= v/v_0 $, and $v_0=2\,\lambda\, k_0$. Neglecting $\mathcal{O}\big (k^2 \big)$ terms yields the linearized Hamiltonian
\begin{align}
	\mathcal{H} \simeq  v_0 \, k \left( \cos\phi \, \sigma_x \, + 
	\sin\phi \, \sigma_y \right) + \Delta \, \sigma_z  \,.   
\end{align}	
The dispersion is given by $\mathcal E = \pm \sqrt{( v_0 \, k)^2 \, + \Delta^2 }$, where
$k^2 \simeq  \frac{\left (k_x^2+k_y^2-k_0^2 \right )^2}  {4\,k_0^2} 
+ \alpha^2 \,k_z^2 $. The SOC acts like a mass term by inducing a gap equal to $2 \Delta$ along each nodal line. 
 
An important feature of the NLSM is that only one component of BC is non-zero, making it impossible to get an MHE response through an orthogonal rotation of the external field or the sample. In particular, one can show that a non-zero BC component
$\Omega_{\theta} = \partial_{k} A_{\phi} - \partial_{\phi} A_{k}
= \frac{\Delta_0 k}  {2 \left ( k^2+ \Delta_0^2 \right )^{3/2}} $ exists only along the angular direction $\hat{\theta}$, where $A_{\mu} 
= i \left \langle u_{+}| \frac{\partial u_{+}}{\partial {\tilde k^{\mu}}} \right \rangle$ is the Berry connection [with $\tilde k^{\mu} \in (k,\phi,\theta)$], $|u_{+} \rangle$ is the normalized eigenstate of the conduction band, and $\Delta_0 =\Delta/ v_0$ represents the scaled SOC strength. This also clearly shows that $\Omega_{\theta}\rightarrow 0$ as $\Delta\rightarrow 0$. Interestingly, we note that $v_{k} = \pm v_0^2 \,k
\,/\, { |\mathcal E | }$, which indicates that $v_{k}>0$ for the upper band [see Fig.~\ref{fig:nodal}(b)]. Equipped with this, we can focus on a positive value of the chemical potential, and compute the components of the conductivity tensors for currents along the polar angle $\phi$. Our results, as demonstrated in Fig.~\ref{fig:nodal}(c), indicate that
$\Delta$ plays a central role in determining the strength of the MHE response. Especially, the MHE responses jump by at least a factor of two as $\Delta$ changes from $0.1$ to $0.2$, indicating that a weaker SOC yields a higher MHE response. This is because, a stronger SOC broadens the band-gap, which reduces the number of available states near the FS, and ends up suppressing the response. We also point out that most of the finite response is seen close to the nodal-line, since at higher energies, the chemical potential cuts the band-edges. Notice that $\sigma^m_{k \phi}$ and $\kappa^m_{k \phi}$ have different profiles when the Fermi level is close to zero energy. This disagreement is expected since both the Wiedemann–Franz law and the Mott relation do not hold in the regime $T=300$ K (considered in the plots) and $\mu$ close to zero.

\section{Discussions and Interpretation of the Results} 
\label{secfinal}

In Sec.~\ref{secresult}, we have investigated the MHE response in various 3D semimetals, showing how different factors (such as symmetry, anisotropy, tilt, and SOC) affect the ballistic conduction, within the Boltzmann transport formalism. A non-zero response is obtained when the non-zero BC of a bandstructure couples to the in-built electric field, and produces a transverse Hall voltage. Being dependent on the transport of charge and heat by itinerant electrons, the MHE conductivity tensors depend on a finite projection of the FS. However, the fate of the response also hinges on the slope of the FS, and how it activates the related component of the BC. If the BC is odd in ${\mathbf k}$, which is true for $\mathcal{T}$-invariant and $\mathcal{I}$-broken materials, then having a symmetric FS exactly cancels the contribution of the left ($-{\mathbf k}$) and right ($+{\mathbf k}$) moving carriers. However, tilting the cone makes the FS asymmetric, thus preventing this exact cancellation. This makes the MHE a useful quantity to probe tilt in topological semimetals.  

Needless to point out, there exist crystal symmetries that can protect the Weyl nodes. One such example is the $n$-fold rotation symmetry around the $z$-axis (usually denoted by $ C_n$), which acts as $\mathcal C_n\, \mathcal{H}(\mathbf k) \,\mathcal C^{-1}_n= \mathcal{H}(R_n \mathbf k)$, where $\mathcal C_n$ is the corresponding rotation operator, and $R_n$ is the $n$-fold rotation matrix in three dimensions acting on 3D vectors. Since the MHE depends on the BC landscape, we are interested in symmetries for which a non-trivial BC emerges, and the energy bands in question give rise to a finite Chern number. A general classification based on $\mathbf k \cdot \mathbf p$ theory~\cite{bernevig_mweyl} shows that while a double-Weyl node in the $xy$-plane can be protected either by a $C_4$ or $C_6$ symmetry, a triple-Weyl node can be stabilized only by a $C_6$ symmetry. Such nodes harbour monopole charges of $\pm 2$ and $\pm 3$, respectively, and thus potentially can show MHE. On the other hand, multi-fold semimetals have been discovered in materials like RhSi, CoSi, and AlPt, which have space group P$2_13$ (number 198) with chiral tetrahedral symmetry~\cite{bradlyn,cano}. In the absence of SOC, such materials contain a three-fold pseudospin-1 node and a two-fold pseudospin-1/2 node, at the points at $\Gamma$ and R of the BZ, respectively. For this situation, the pseudospin-1/2 quasiparticles cannot contribute to MHE (since tilt is forbidden due to symmetry), and any MHE response should come predominantly from the pseudospin-1 quasiparticles. 
For nodal-line semimetals, examples include $\mathrm{Cu_3PdN}$ with space group symmetry Pm$\bar 3$m (number 221), and $\mathrm{CaP_3}$ with space group P-$1$, hosting a nodal line around R-point and near the $\Gamma$-Y-S plane, respectively~\cite{hongming}.

Most semimetallic phases occur at low-symmetry points of the BZ, where anisotropy is generic. Due to their space-group symmetries, the anisotropic multi-Weyl dispersions are either quadratic or cubic in the $xy$-plane, and linear along the $z$-axis. This anisotropy is reflected in the FS and the relevant component of the BC, which makes the components of the Hall conductivity tensor unequal. For instance, a contrasting MHE response can be seen in our result when the Fermi level is close to the node and the tilt is sub-critical. We find that the $xy$-component of MHE response stays constant, while the $zx$-component rises sharply. Changing the non-linear dispersion from quadratic ($J=2$) to cubic ($J=3$) simply increases the strength of the response, since the components of the BC are proportional to either $J$ or $J^2$. Going beyond the critical point also increases the response, as electron-hole pockets form at the FS. Based on this, we expect the MHE to be weak in $\mathcal{T}$-invariant $\mathrm{TaAs}$ class of materials possessing type-I Weyl quasiparticles, while type-II semimetals like $\mathrm{WTe}_2$, $\mathrm{MoTe}_2$, $\mathrm{WP}_2$, and $\mathrm{Ta_3S_2}$ should have pronounced MHE effects~\cite{gao}.

Recently-discovered chiral materials like $\mathrm{RhSi}$ and $\mathrm{CoSi}$, representing the multi-fold semimetals, can also display MHE. The overall trends of the MHE response in these systems, and the dependence on tilt, are reminiscent of the $\mathcal{T}$-invariant Weyl cones, as both share similar FS and BC behaviour. The only difference is that multi-fold semimetals have more bands near the band-crossing point, that can participate in the conduction in the type II phase -- this makes the MHE response stronger than its two-band Weyl counterparts in this regime.

We have also shown that the requirement of tilt is not always necessary (see Sec.~\ref{sec_sweyl}). This is true for $\mathcal T$-broken materials where BC is either positive or negative throughout the BZ, and thus symmetric cancellations cannot occur. Interestingly, our result shows that increasing the distance between the Weyl nodes in momentum space amplifies the MHE. Materials like $\mathrm{WP_2}$ and $ \mathrm{MoP_2}$ have a large separation of their Weyl nodes, and are thus expected to show a strong MHE response. 
However, the absence of $\mathcal{T}$ leads to non-zero AHE at the same time, and the anomalous Hall conductivity for Weyl semimetal is also proportional to this separation between the Weyl nodes~\cite{ashvin}. Hence, the tricky part is that the AHE contributions can mask the MHE signatures. A more realistic calculation of different Hall contributions within a lattice framework will be more illuminating in this regard. 

\section{Summary and Outlook}
\label{conclude}

2D Hall effect is well-known for a long time, and is now a routine technique in condensed matter experiments for identifying various characteristics of novel materials. Over the last few decades, several attempts have been made to extend the idea of 2D quantum Hall effect (QHE) to 3D~\cite{druist, stromer}. One of the widely-used methods is stacking multiple weakly coupled layers to form a quasi-2D platform. But due to weak interlayer coupling, the FS of such quasi-2D platforms is not exactly similar to a 3D FS, and this is a fundamental challenge in material fabrication~\cite{li_3dhall}. More recent attempts to realize 3D QHE have been made by using materials like $\mathrm{ZrTe_5}$~\cite{liang_zrte5} and $\mathrm{HfTe_5}$~\cite{galeski_hfte5}. Apart from this, a non-linear Hall effect has been studied, where the double-frequency Hall signal is measured in response to an excitation current, in materials like  $\mathrm{TaIrTe_4}$~\cite{kumar}, $\mathrm{Cd_3As_2}$~\cite{shvetsov}, and $\mathrm{Bi_2Se_3}$~\cite{hyunsoo}. Given these rapid developments, we are hopeful that in the near-future, experimental realizations of the 3D MHE will be perfected soon. Hence, this work is extremely timely in providing the theoretical predictions for such experiments.

While determining the MHE response, it is important to remember two important facts:
\\\textbf{(1)} the MHE depends on $\Omega_c (\mathbf{k})$ with $v_a (\mathbf{k})>0$. If the FS is symmetric, the states participating in transport have $\Omega_c$-values which are exactly equal and opposite in sign, leading to a complete nullification. This symmetric situation can be avoided if a tilt along the $\hat{a}$-axis exists, which makes the relevant FS projections asymmetric. Thus, we have considered separate tilts along the $z$-axis and the $y$-axis, while calculating the $xy$- and $zx$-components of the response, respectively, for the anisotropic multi-Weyl semimetals. A similar argument follows for the $yz$-component, necessitating a tilt along the $x$-axis.
\\\textbf{(2)} 
Weyl and multi-Weyl semimetals feature nodes (or valleys) with opposite chiralities.
Since MHE is proportional to chirality and tilt, there might be valley cancellations, as pointed out in Ref.~\cite{das}. Hence, a non-zero contribution is possible only if the tilts are opposite in the two valleys, or if the corresponding Fermi levels lie at different energies (as seen for the case of the multi-fold semimetals). 
\\We also note that in most Weyl materials, such as the $\mathrm{TaAs}$ family class, the Fermi level is either at the band-degeneracy point, or remains very close to it (cf. Ref.~\cite{weyl-rev}). This limits the higher-band influence for these materials. However, higher bands will play a significant role in multi-fold semimetals, when the Fermi level is tuned to higher energies.

For the TRS-broken case, BC peaks for long-wavelengths around $k_x=0$ axis [cf. Fig.~\ref{fig:weyl}(b)]. With a chemical potential close to zero, FS cuts a region with a large amplitude of the relevant BC component. This explains why the conductivity tensor component peaks at moderate doping. Away from this zero-doping region, although the FS is larger, it mostly activates a low amplitude of the BC, which leads to a drop in the response. We have found that the response is proportional to the distance between the Weyl nodes in the momentum space. The experimental realization for TRS-broken semimetal might be possible using periodic stacking of normal and topological insulators~\cite{burkov}, such that a Zeeman field breaks $\mathcal T$. Recently-discovered magnetic Weyl semimetals~\cite{ning,destraz} are also promising platforms, given that they break $\mathcal T$, and host large intrinsic BC values.

In $\mathcal T$-invariant but $\mathcal I$-broken materials, BC satisfies $\mathbf{\Omega(\mathbf k)} =- \mathbf{\Omega(-\mathbf k)}$, and therefore the signs of the BC are opposite for two sets of carriers with opposite momenta. This leads to a perfect cancellation in the absence of factors such as tilt or strain (see the discussion above). Hence, an asymmetry in the FS is necessary to generate a finite Magnus Hall response. This makes the MHE a possible probe for FS asymmetry in TRS-invariant systems, and should be contrasted with the response expected for TRS-broken materials. We also note that the MHE conductivity in the $ab$-plane depends on the component $\Omega_c$ (where $c$-axis is perpendicular to the plane), with $v_a>0$. This indicates in turn that a single FS projection along the $ac$-plane will determine the transport. TRS-preserving multi-Weyl semimetals show different values of response for different planes, because of anisotropic dispersions, and this can be further tuned through the tilt parameter. Since BC values peak for small ${\mathbf k}$, intuitively one would expect large values of conductance for small Fermi energies. This is true for the $zx$-component. Whereas, for the $xy$-component, FS activates carriers with opposite values of the BC components, which results in an internal cancellation, and the MHE remains constant near zero-doping. However, if the cones are over-tilted, we find that the MHE response is amplified, since more states are now accessible to the FS.

Multi-fold semimetals have different FS structures than the multi-Weyl semimetals discussed above. As shown in Fig.~\ref{fig:multifold}(b), the FS is a small, closed curve for small tilt and moderate doping. The MHE response is mostly constant and low in this regime, because active states near the FS correspond to BC components with opposite signs. However, beyond a critical tilt, the FS becomes open, and accordingly, we observe an increase in the response.

The above discussions show that one of our main results involves the important role of the tilt for nodal-point semimetals, because a non-zero MHE response requires a finite tilt along a specific coordinate-axis (that is dictated by the relevant component of the BC appearing in the expressions for the corresponding transport coefficient). The tilt prevents the symmetric cancellations of the activated BC weights over the FS. Consequently, such tilt-dependent MHE response is a distinguishing feature between semimetals with and without tilt.
For example, the peak of $\sigma_{k\phi}^m$ increases if $\Delta$ is increased, and at moderate doping, a higher SOC yields a larger MHE response. This is because, MHE coefficients are proportional to the magnitude of $\Omega_{\theta}$, which increases with SOC.

Unlike the TRS-broken Weyl systems, the BC of nodal-line semimetals is positive everywhere. This makes it possible to have a finite MHE without the need for asymmetry (such as tilt or strain). It is simple to understand that the relevant BC component peaks at small momenta. Thus, MHE conductances have a higher yield near zero-doping, as the FS activates a large value of the BC amplitude. The magnitude of BC drops off rapidly away from ${\mathbf k}=0$, which explains smaller MHE responses away from zero chemical potential.

\begin{table}
\begin{tabular}{|c|c|c|c|}
\hline
\textbf{\makecell{Behaviour of $\sigma^m_{ab}$\\near $\mu=0$}}
&\textbf{Remarks}&\textbf{Possible material properties}  
&\textbf{
\makecell{
Further characterization}}\\
\hline
Close to zero & 
\makecell{Complete cancellation
of active\\states or zero BC component} &
\makecell{Topologically trivial / \\
$\mathcal T$-invariant with no tilt / \\
NLSM without SOC}
& -- \\
\hline
Almost constant &
\makecell{Partial cancellation of\\
MHE contributions}&
$\mathcal T$-invariant with small tilt &
\makecell{If rotating the sample changes the response,\\then multi-Weyl,
otherwise Weyl or multi-fold}\\
\hline
\makecell{Monotonically\\increasing} &
\makecell{Positive velocity carriers
\\have a net BC contribution}
&
\makecell{
$\mathcal T$-broken semimetal / \\
$\mathcal T$-invariant with tilt / 
\\NLSM with SOC
} 
& 
\makecell{
The peak magnitude depends on\\
(1) separation of nodes in momentum space\\
 for $\mathcal T$-broken Weyl\\
(2) strength of SOC for NLSM with SOC
}\\
\hline
\end{tabular}
\caption{\label{table:1}
Characterization of 3D topological semimetals based on the Magnus Hall response near zero chemical potential.} 
\end{table}

From the above discussions, it is evident that the MHE responses show noticeable changes mostly near zero chemical potential. In particular, the electrical conductivity can be zero, constant, or monotonically increasing in this regime, depending on the underlying symmetries of the semimetal under consideration. This will enable us to make useful predictions about those symmetries based on the MHE response curves obtained from experiments. We summarize such observations in Table~\ref{table:1}.

We have found that increasing the momentum-space separation of the two nodes with opposite chiralities amplifies the MHE response. However, as pointed out earlier, measuring the actual MHE response using this property can be tricky, as AHE is also known to increase with such increasing node separations. Another important yet previously overlooked part, that we have focussed on in our paper, is the calculation of different components of the MHE conductivity. Ideally, Weyl cones are isotropic at low energies. But in reality, they can have quadratic and cubic anisotropy, as seen in multi-Weyl semimetals. Such anisotropy can massively impact different MHE components to the point where one component is preferred over another in experiments, simply because the former yields more response. This can be seen in the upper panel of Fig.~\ref{fig:mweyl}(c), where the $zx$-component is an order of magnitude larger than the $xy$-component, in the vicinity of $\mu=0$.

The intrinsic contribution to the MHE, which is a non-linear Hall effect, has a geometric nature because of its connection to the BC. Consequently, it has opened up heretofore unexplored and promising prospects for future research in the area of topological phases, by driving the study of quantum transport and topological physics to the non-linear response regime.
The fact that MHE requires no external magnetic field can be of great importance, as it potentially paves the pathway for fabricating new types of devices based on the non-linear Hall effects. Some examples of current applications include (1) few-layer $\mathrm{WTe_2}$ being used as a reading mechanism of the BC memory~\cite{shao}, (2) strain sensors for designing piezoelectric-like devices~\cite{jiang}, and (3) current-rectification without semiconductor junctions~\cite{hiroki}. 

Our analysis, considering various 3D semimetals, shows that the MHE conductivity tensors can be used as a complementary platform to map out various properties of topological semimetals, which can be used in conjunction with other probes \cite{ipsita-sajid,ips-kush,
ips_qbt_tunnel,*ips3by2,*ipsita-aritra,*ipsfloquet}. One interesting direction will be to compute the effects of disorder \cite{emil,*emil2,rahul-sid,*ipsita-rahul,*ips-birefringent} and / or interactions \cite{roshtami,kozii,Mandal_2020}. However, a major impediment in these directions is that no consistent theory exists beyond the semiclassical approximation, which breaks down when quantum effects are robust.
A possible way to make progress is to apply the invariant measure approach, which is
currently being developed for simple transport properties \cite{ips-klaus}.

\section*{Acknowledgments}

We thank Shivam Yadav for carefully going through the manuscript, and providing valuable feedback.
The research is partially funded by the National Science Centre (Narodowe Centrum Nauki), Poland, under the scheme PRELUDIUM BIS-2 (grant number 2020/39/O/ST3/00973).

\bibliography{ref.bib}

\end{document}